# A Quantum Yield Map for Synthetic Eumelanin

Stephen Nighswander-Rempel,[1]* Jennifer Riesz,[1,2], Joel Gilmore,[3] and Paul Meredith[1,2]

[1] Centre for Biophotonics and Laser Science
[2] Soft Condensed Matter Physics Group
[3] Soft Condensed Matter Theory Group
School of Physical Sciences, University of Queensland
St. Lucia, QLD, Australia 4067
Email: snighrem@physics.uq.edu.au

**Abstract**
The quantum yield of synthetic eumelanin is known to be extremely low and it has recently been reported to be dependent on excitation wavelength. In this paper, we present quantum yield as a function of excitation wavelength between 250 and 500 nm, showing it to be a factor of 4 higher at 250 nm than at 500 nm. In addition, we present a definitive map of the steady-state fluorescence as a function of excitation and emission wavelengths, and significantly, a three-dimensional map of the "specific quantum yield": the fraction of photons absorbed at each wavelength that are subsequently radiated at each emission wavelength. This map contains clear features, which we attribute to certain structural models, and shows that radiative emission and specific quantum yield are negligible at emission wavelengths outside the range of 585 and 385 nm (2.2 and 3.2 eV), regardless of excitation wavelength. This information is important in the context of understanding melanin biofunctionality, and the quantum molecular biophysics therein.

**Introduction**

Eumelanin is a biological pigment found in many species (including humans). In the human body, it is known to act as a photoprotectant in the skin and eyes,[1] a function derived both from its strong absorption throughout the UV and visible wavelengths (Fig. 1) and from its low quantum yield. Paradoxically, eumelanin precursors have also been implicated as photosensitisers leading to the development of melanoma skin cancer.[2] The chemical properties of eumelanin are therefore a topic of intense scientific interest;[3,4] in particular, an understanding of the radiative and non-radiative de-excitation processes of eumelanin is critical to unlocking its role with respect to melanoma, and understanding these processes is a key goal of many groups, including ours.

Eumelanin fluorescence has been extensively studied over the past three decades; however, much of the reported literature is inconclusive and inconsistent.[5] This is in part due to the inner filter effect (attenuation of the incident beam) and strong reabsorption of the emission, even at very low concentrations. Since eumelanin absorbs very strongly and its absorbance profile is exponential in nature, these effects distort the spectra significantly. However, a correction method has recently been used to successfully recover the actual eumelanin emission[6] and excitation spectra[7] at several excitation and emission wavelengths respectively. The first study clearly showed that eumelanin emission is dependent upon excitation wavelength, as increases in excitation wavelength red-shifted the emission peak and reduced its intensity. In order to determine the limits of this effect, we report here a significant extension of that preliminary study: a complete set of emission and excitation spectra for synthetic eumelanin at 1 nm intervals over the entire visible and UV range. We believe this to be the most complete study of the steady-state fluorescence of eumelanin to date. As such, we hope it can act as a reference point for further spectroscopic studies.

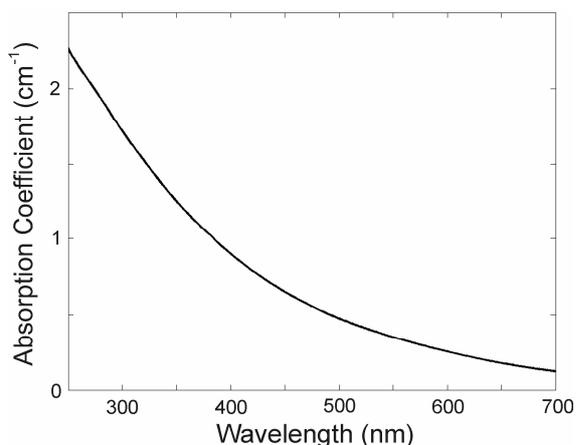

Fig. 1. Absorbance spectrum of synthetic eumelanin solution in pH 10 NaOH.

Further to this, we provide a complete description of the dependence of the radiative quantum yield on excitation wavelength. The quantum yield of eumelanin has been shown to be 18% lower for 410 nm excitation than for 350 nm excitation;[8] this wavelength-dependence is an unusual characteristic among fluorophores. The current study provides quantum yield values for all excitation wavelengths between 250 and 500 nm. In this effort, we introduce a quantity that we call the "specific quantum yield" for eumelanin. This is the fraction of photons absorbed at a specific excitation wavelength that are emitted *at a specific emission wavelength* and can be depicted for all excitation and emission wavelengths in a three-dimensional "quantum yield map". Note that the traditional quantum yield is given by the integral of the specific quantum yield over emission wavelengths. For a molecule with complex energy dissipation processes and broad spectroscopic features such as eumelanin, the specific quantum yield is a valuable parameter for spectroscopic analysis. Even for compounds whose quantum yield exhibits no wavelength-dependence, the specific quantum yield is valuable for spectroscopic studies as it represents the emission distribution normalised to absorption. In addition to reporting the specific quantum yield map for eumelanin, we present the general method for determination of the specific quantum yield for any compound.

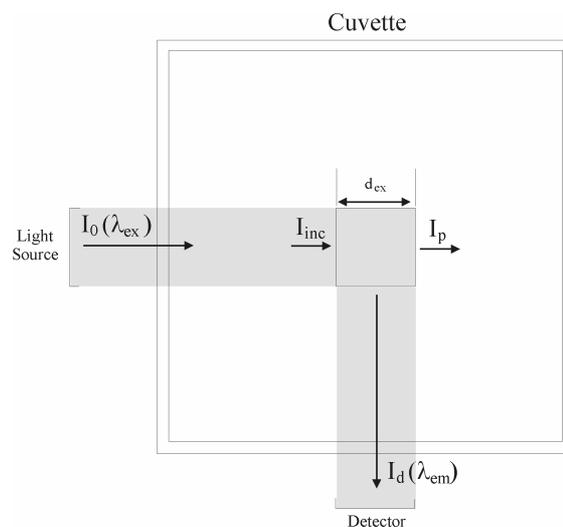

Fig. 2. Diagram of cuvette and excitation volume, with respect to excitation and emission beam width.

**Experimental Section**
Computational Methods
We seek to determine the relationship between the specific quantum yield values, as defined above, and the excitation and emission wavelengths ($\lambda_{ex}$ and $\lambda_{em}$, respectively). If we take as a typical measurement geometry the configuration shown in Fig. 2, then a small volume is defined in the centre of the cuvette by the slit widths for the incoming and outgoing beams. This is the volume from which fluorescence is detected, given the instrumental design. We define:

$N_a(\lambda_{ex})$ = Total number of photons absorbed in the central volume
$N_e(\lambda_{ex}, \lambda_{em})$ = Total number of photons emitted from the central volume

The specific quantum yield, as a function of $\lambda_{ex}$ and $\lambda_{em}$ is then defined as:

$$Q(\lambda_{ex}, \lambda_{em}) = N_e(\lambda_{ex}, \lambda_{em}) / N_a(\lambda_{ex}) \qquad 1$$

$N_a$ is the difference between the number of photons incident on the central volume and the number of photons remaining after passing through the volume. Since the number of photons is directly proportional to the light intensity with some proportionality constant $K$, we have

$$N_a(\lambda_{ex}) = K[I_{inc}(\lambda_{ex}) - I_p(\lambda_{ex})] \qquad 2$$

Moreover, by the Beer-Lambert law,

$$I_p(\lambda_{ex}) = I_{inc}(\lambda_{ex}) e^{-\alpha(\lambda_{ex}) d_{ex}} \qquad 3$$

where $\alpha(\lambda_{ex})$ is the absorption coefficient of the sample at $\lambda_{ex}$ and $d_{ex}$ is the width of the central volume. Combining equations 2 and 3 yields

$$N_a = K I_{inc}(\lambda_{ex})[1 - e^{-\alpha(\lambda_{ex}) d_{ex}}] \qquad 4$$

Consider now the photons that are emitted from the excitation volume. If we define $I_e$ to be the total intensity emitted from the excitation volume (in all directions), then $I_d$ (that fraction of $I_e$ that is detected) will be proportional to $I_e$. The proportionality constant $C$ (as defined in Eq. 5) will be less than one and dependent only on the system geometry and the detector sensitivity, not on $\lambda_{ex}$ or $\lambda_{em}$. Thus,

$$N_e(\lambda_{ex}, \lambda_{em}) = K I_d(\lambda_{ex}, \lambda_{em}) / C \qquad 5$$

The specific quantum yield is then given by (combining Eqs. 1, 4 and 5):

$$Q(\lambda_{ex}, \lambda_{em}) = \frac{I_d^*(\lambda_{ex}, \lambda_{em})}{C(1 - e^{-\alpha(\lambda_{ex}) d_{ex}})} \qquad 6$$

Here, $I_d/I_{inc}$ has been replaced with $I_d^*$, reflecting the fact that raw emission intensity data recorded by the spectrometer will have been pre-corrected for variations in lamp intensity. Also, in order to account for probe attenuation and emission reabsorption within the sample (which are significant not only for eumelanin but also for common quantum yield standards such as quinine[9]), a correction has been applied to the raw spectra,[8] prior to determination of the quantum yield. The value typically reported as the quantum yield (the 'traditional' quantum yield, $\phi$) will then be the integral of Eq. 6 over all emission wavelengths:

$$\phi(\lambda_{ex}) = \frac{1}{C} \frac{\int I_d^*(\lambda_{ex}, \lambda_{em}) d\lambda_{em}}{1 - e^{-\alpha(\lambda_{ex}) d_{ex}}}, \qquad 7$$

Note that the factor $1/C$ is a normalising parameter dependent only on the system geometry and the detector sensitivity. In order to determine this factor, we can measure the absorbance and emission spectra of a standard with a known quantum yield $\phi_{st}$. Then a simple rearrangement of Eq. 7 yields

$$\frac{1}{C} = \frac{\phi_{st}(1 - e^{-\alpha_{st}(\lambda_{ex}) d_{ex}})}{\int I_{d,st}^*(\lambda_{ex}, \lambda_{em}) d\lambda_{em}} \qquad 8$$

The above equations for the quantum yield are equivalent to standard methods provided in the literature.[10] Note that typically, the ratio of integrated emission to absorption coefficient α is used, whereas the present discussion uses the ratio of integrated emission to $1-e^{-\alpha d}$. The former ratio is based upon the approximation $e^{-\alpha d}=1-\alpha d$, which is not valid for studies such as ours in which the sample (melanin) has large absorption coefficient values. For more precise results, we have measured the absorbance and emission of the standard solution for several different concentrations and plotted the expression in Eq. 8 for thirty concentrations (Fig. 3). $1/C$ is then given by the gradient of a linear regression.

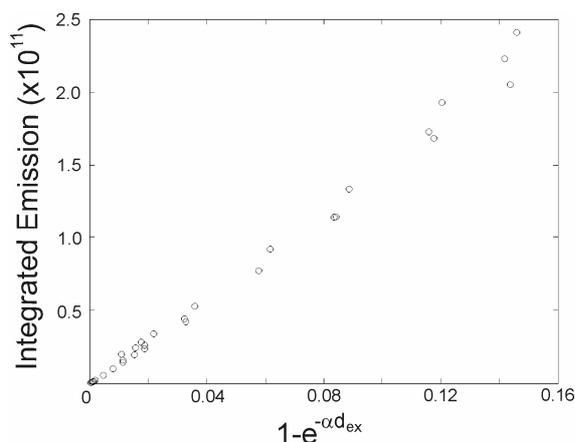

Fig. 3. Integrated fluorescence emission as a function of the absorbance ($\alpha$) for 30 quinine sulphate solutions ($1\times10^{-6}$ to $1\times10^{-4}$ M in 1 N $H_2SO_4$). The length of the excitation volume ($d_{ex}$) was assumed to be 0.1 cm.

Sample Preparation
Synthetic eumelanin derived by the nonenzymatic oxidation of tyrosine was purchased from Sigma-Aldrich (Sydney, Australia) and was treated by acid precipitation in order to remove small molecular weight components, following the method of Felix et al.[11] Briefly, dopamelanin (0.0020 g) was mixed in 40 mL high-purity 18.2 MΩ MilliQ deionized water and 0.5 M hydrochloric acid was added to bring the pH to 2. Solutions were centrifuged and the black precipitates were repeatedly washed in 0.01 M hydrochloric acid and then deionized water. A 0.0050% solution (by weight) of the remaining precipitate was prepared in de-ionised water. This concentration was selected to maximise fluorescence, while minimising re-absorption and inner filter effects (the correction for these effects has been shown to be effective at this concentration; at higher concentrations, scattering effects reduce the accuracy of the correction[8]). To aid solubility, the solution was adjusted to pH 10 using NaOH (as in previous studies[6,8]). Given that high pH also enhances polymerization,[12] this adjustment also ensured that the presence of any residual monomers or small oligomers in the solution was minimised. A pale brown, apparently continuous dispersion was produced. Quinine sulphate (Sigma-Aldrich) was used without further purification at thirty different concentrations ($1\times10^{-6}$ M to $1\times10^{-4}$ M in 1 N $H_2SO_4$ solution) as a standard for the determination of the radiative quantum yield.

Spectroscopy
Absorbance spectra were recorded using a Perkin Elmer (Norwalk, CT) Lambda 40 spectrophotometer with a 240 nm/min scan speed and 2 nm bandpass. All spectra were collected using a 1 cm square quartz cuvette. Solvent scans (obtained under identical conditions) were used for background correction.

Fluorescence emission spectra for eumelanin and quinine sulphate were recorded using a Jobin Yvon (Edison, NJ) Fluoromax 3 fluorimeter with a 3 nm bandpass and an integration time of 0.5 s. Matrix scanning software allowed excitation and emission intervals of 1 nm. Solvent scans were again performed under identical instrumental conditions for background correction. Spectra were pre-corrected to account for differences in pump beam power at different excitation wavelengths using a reference beam. All emission spectra were corrected for reabsorption and inner filter effects using the method outlined previously.[8] Quantum yields were calculated using the method outlined above with standard values.[9] Since the quantum yield of quinine is temperature-dependent, the ambient temperature surrounding the cuvette was measured to be 35°C, resulting in a 2.5% shift from the published value of 0.546.

**Results and Discussion**

The fluorescence map for synthetic eumelanin is shown as a function of two variables in Fig. 4a as both a three-dimensional projection and a contour map. The first- and second-order Rayleigh peaks have been removed from the spectra manually, and the first- and second-order Raman bands were removed by background subtraction in the correction procedure (which accounted for probe attenuation and emission reabsorption).

Excitation and emission spectra extracted from these maps correspond well with spectra reported previously.[7,8] It is clear from this map that while emission spectra (vertical cross-sections of the map) exhibit only a single peak, excitation spectra (horizontal cross-sections of the map) present multiple peaks, particularly for emission wavelengths between 450 and 550 nm. This is due to the fact that regardless of excitation wavelength, virtually all emission occurs at wavelengths between 385 nm (3.2 eV) and 585 nm (2.2 eV). The lack of emission beyond 600 nm has been previously reported for excitation wavelengths between 360 and 380 nm as a low energy tail in the emission spectra that is constant with excitation energy (in shape and magnitude).[8] The present data show that this feature is maintained for a much broader range of excitation wavelengths. The map and its extracted excitation spectra also reveal the following interesting features:

- High emission at all energies between 2.1 and 4.8 eV (600 and 260 nm) when excited at energies greater than 4.8 eV (wavelengths shorter than 260 nm).
- A strong emission maximum at 2.7 eV (460 nm) for excitation at all energies greater than 4.2 eV (shorter than 295 nm; the exact wavelength is off the map).
- A broad band in emission spectra shifting from 435 nm to 517 nm (2.9 to 2.4 eV) as the excitation wavelength is increased from 310 to 460 nm (4.0 to 2.7 eV; Fig. 4b).
- Greater emission at all wavelengths between 450 and 590 nm (2.8 and 2.1 eV) when excited at 365 nm (3.40 eV) than when excited at higher or lower wavelengths (Fig. 4c).
- Greater emission at all wavelengths between 500 and 540 nm (2.5 and 2.3 eV) when excited at 490 nm (2.54 eV) than when excited at higher or lower wavelengths (Fig. 4c).

While eumelanin has long been considered to be a heteropolymer of indolic units[13] and is still often cited as such,[14] recent studies have supported an alternate model for the secondary structure of eumelanin as a collection of oligomers of varying size that may or may not be stacked.[15] According to this model, each of these oligomers have a slightly different HOMO-LUMO (highest occupied molecular orbital – lowest unoccupied molecular orbital) gap energy and the broad spectral features of eumelanin may actually be the result of superpositioning of many different, narrower peaks, each corresponding to a slightly different chemical species. We refer to this as the chemical disorder model.

Within this model, existence of an emission peak in Fig. 4a may reflect either increased concentrations of the corresponding chemical species, increased molar absorptivity at the excitation wavelength, or increased quantum yield of that species. These peaks can not be due solely to differences in concentration or molar absorptivity, since such differences would result in corresponding peaks in the absorbance spectrum of eumelanin. Moreover, while greater absorption intuitively leads to greater fluorescence, and eumelanin absorbance increases monotonically towards higher energies (Fig. 1), Fig. 4 clearly demonstrates that this does not lead to monotonic increase in fluorescence. Therefore, these data clearly indicate that the quantum yield is dependent on excitation-wavelength .

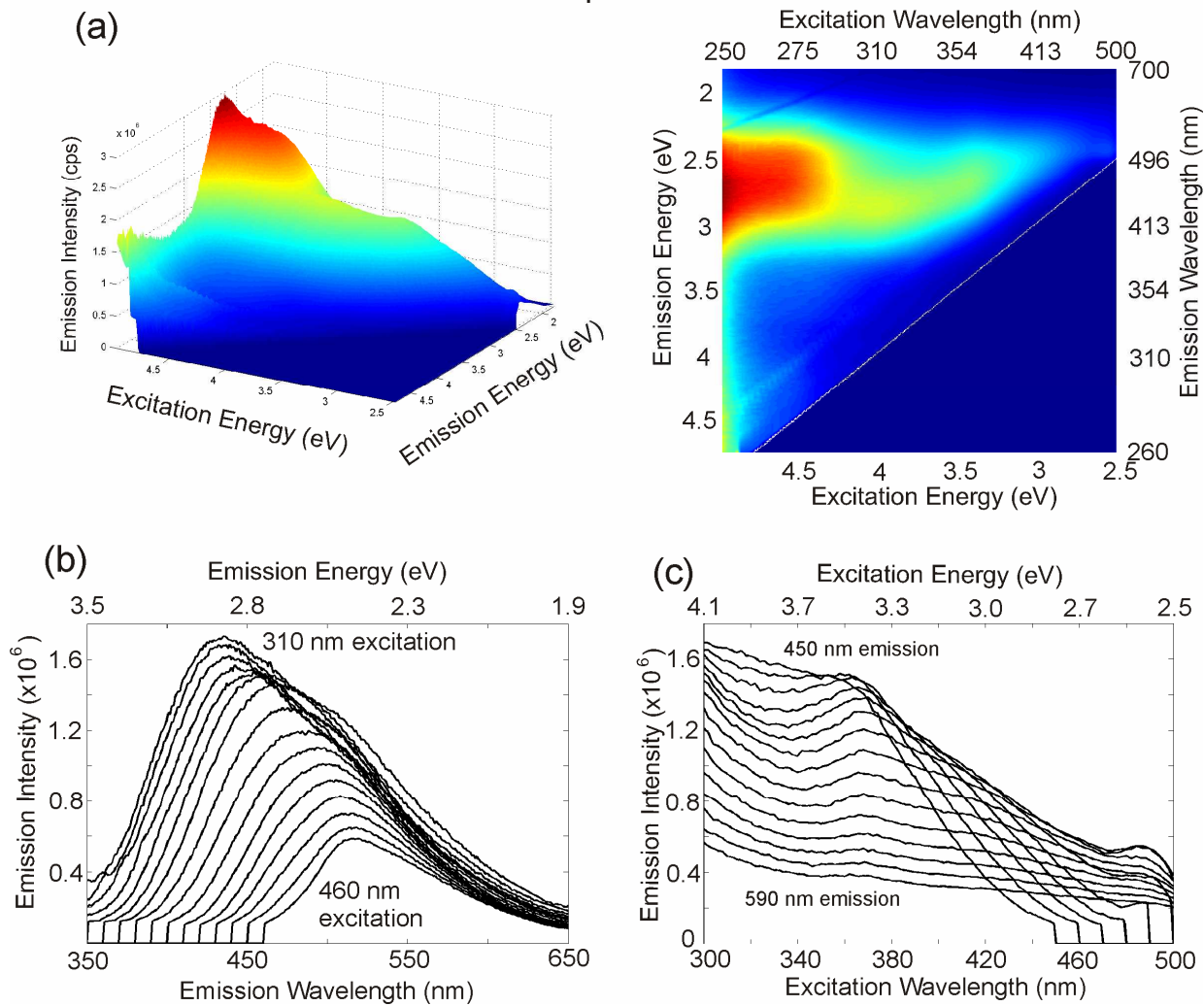

Fig. 4. (a) Reabsorption-corrected fluorescence map for synthetic eumelanin. High emission intensity is shown in red and low intensity is shown in blue. (b) Excitation spectra extracted from fluorescence map for emission wavelengths between 360 and 450 nm show a broad band shifting in position between 280 and 320 nm. (c) Excitation spectra for emission wavelengths between 450 and 590 nm show a peak constant in position at 370 nm but varying in intensity.

Calculation of the specific quantum yield as discussed above removes the effects of concentration and molar absorptivity. Specific quantum yield values shown in Fig.

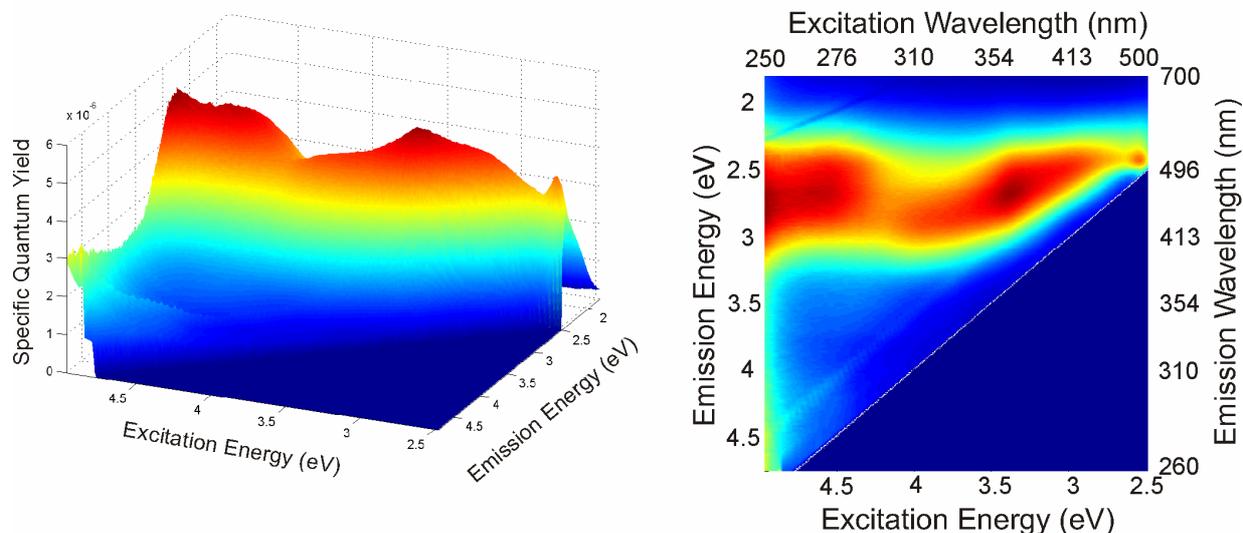

Fig. 5. Specific quantum yield map: the fraction of photons absorbed at each excitation wavelength that are emitted at each emission wavelength. Two peaks are evident and limiting values at high- and low-emission are observed.

5 can also be interpreted as the probability that an absorbed photon will be re-emitted at a particular wavelength. Note that all the values shown are extremely small, the maximum being $5.2 \times 10^{-6}$ (0.00052%), and this is representative of the fact that eumelanin has strong electron-phonon coupling, allowing it to de-excite non-radiatively, making it an excellent photoprotectant.

All of the features listed above for the fluorescence map are retained in this quantum yield map, and some (especially those at higher excitation wavelengths) have become much more prominent. Of particular note is the presence of three strong peaks of comparable intensity at excitation wavelengths of 3.4 eV, 4.6 eV and at some energy above 5 eV. Since higher quantum yield values reflect greater radiative decay and hence weaker electron-phonon coupling, these bands suggest that the corresponding chemical species are more loosely connected to the melanin compounds. Since acid precipitation makes the presence of monomers and small oligomers unlikely, they may reflect monomers singly-linked to a larger oligomeric structure or terminal elements in a chain. Along these lines, we have previously suggested that the 3.4 eV peak in excitation spectra is due to DHICA (5,6-dihydroxyindole-2-carboxylic acid) due to the peak's proximity to absorption peaks in the DHICA monomer.[7,16] If this is the case, then the peak near 5 eV may be due to its higher-order transition.

These data are supported by calculation of the traditional quantum yield for eumelanin, which is shown to agree excellently with the very low values (on the order of $10^{-4}$) previously reported (Fig. 6).[8] The fact that the traditional quantum yield at 3.4 eV (as well as the specific quantum yield) is greater than for higher or lower excitation energies supports the suggestion that this chemical species is weakly linked to the larger structure. On the other hand, the peak at 4.6 eV is less clearly distinguished. The reduced indolequinone monomer (IQ; tautomer 3b in Ref. [17]) has been shown to exhibit absorption bands at 5.2 eV and 2.6 eV. Since the higher-order transition

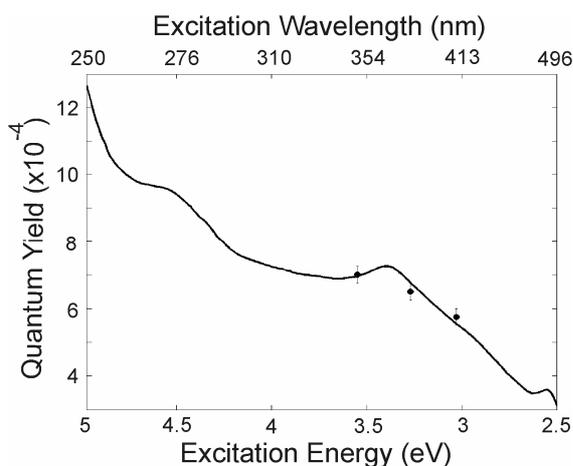

Fig. 6. (a) The traditional quantum yield as a function of excitation energy across UV and visible wavelengths (solid line). Circles and error bars show previously measured quantum yield values (taken from Ref. [8])

involves more delocalised molecular orbitals, they might be expected to redshift more with polymerisation than the lower-order transition. Given this, these data fit well with the 4.6 eV peak and the weaker 2.6 eV peak presented here, suggesting that both IQ and DHICA monomers may be present at the edges of the melanin oligomers. An alternative explanation of these features involves the half-reduced semi-quinone (SQ). SQ hexamers exhibit an absorption peak near 480 nm (2.6 eV).[4] Since a different IQ tautomer (3a in Ref. [17]) has a peak at 5.0 eV without any peak near 2.6 eV, the spectral features described here may reflect contributions by IQ (tautomer 3a in Ref. [17]) and DHICA on the periphery of large SQ oligomers.

The broad emission band shown in Fig. 4b is even more pronounced in Fig. 5. Within the chemical disorder model, this feature can be explained by selective excitation of different sized oligomers. Simulations of larger oligomeric structures show that polymerisation leads to progressive red shifting of the gap[18] and increased delocalisation of the electronic wavefunctions. Stark et al. in their latest paper have augmented these findings, demonstrating further red shifting with stacking.[4] Thus, excitation with lower-energy light can be expected to excite larger oligomeric structures, resulting in lower-energy emissions as well.

With this consideration, the decline in quantum yield between 3.4 and 2.6 eV (Fig. 6) may simply reflect a limit to the size of melanin oligomers and a corresponding decrease in the number of possible transitions along which it may de-excite. As with the fluorescence map, the specific quantum yield is almost negligible for all emission energies lower than 2.2 eV (wavelengths longer than 585 nm) and greater than 3.2 eV (shorter than 385 nm), regardless of excitation energy (Fig. 5). This lack of fluorescence suggests that de-excitation along transitions with energy gaps outside this range is much less likely than de-excitation within this range. Note that this does not indicate a lack of possible transitions at those energies, since there is significant absorption at all wavelengths shorter than 385 nm.

For most substances, the traditional quantum yield is constant with excitation wavelength (and is usually quoted as a single value). This is clearly not the case for eumelanin; the yield varies by a factor of 4 over the range 250 nm to 500 nm. Our method for determining the traditional yield has allowed us to plot it as a function of the excitation energy, fully characterising the yield over the UV and visible range, and suggest possible chemical interepretations of these data. By understanding the chemical structure of eumelanin and its corresponding decay pathways, we hope to understand both how it interacts with its biochemical environment *in vivo* and maybe even its specific roles in the development of melanoma skin cancer.

## Conclusions

We have presented the most comprehensive study to date of steady-state eumelanin fluorescence for UV and visible wavelengths, correcting for attenuation and reabsorption effects. These data demonstrate upper and lower bounds on emission that are independent of excitation wavelength between 250 and 500 nm. Moreover, we have introduced a new parameter, the "specific quantum yield", which characterises the radiative (and non-radiative) decay properties of eumelanin more completely than emission or excitation spectra alone. Finally, we demonstrate that the traditional quantum yield is extremely low (on the order of $10^{-4}$) and highly dependent on wavelength, increasing by a factor of 4 with excitation energies between 2.5 and 5 eV (500 and 250 nm). We hope that these data will serve as a reference point for further spectroscopic studies of these fascinating and important biomolecular systems.